# Probing electronic transitions and defect-induced Urbach tail bands in functional perovskite oxides using diffuse reflectance spectroscopy and the Kubelka-Munk function

Ramachandran Balakrishnan[1,a)], Priyambada Sahoo[2], Balamurugan Karuppannan[3,4], and Ambesh Dixit[2]

AFFILIATIONS

[1]Department of Physics, Sethu Institute Technology, Pulloor, Kariapatti-626115, Virudhunagar, Tamil Nadu, India.

[2]Advanced Materials and Devices Laboratory, Department of Physics, Indian Institute of Technology Jodhpur, Jodhpur 342011, Rajasthan, India

[3]Department of Physics, School of Engineering and Technology, Dhanalakshmi Srinivasan University, Samayapuram, Tiruchirappalli-621112, Tamil Nadu, India.

[4]Center for Research, Dhanalakshmi Srinivasan University, Samayapuram, Tiruchirappalli-621112, Tamil Nadu, India.

[a)]Author to whom correspondence should be addressed: ramskovil@gmail.com

ABSTRACT

We conducted a detailed study of electronic transitions and defects-induced Urbach tail bands in various functional perovskite oxides ($V_2O_5$, $BaSnO_3$, $PbZr_{0.52}Ti_{0.48}O_3$, $BiMnO_3$, and $BiFeO_3$) using diffuse reflectance spectroscopy (DRS). We analyzed their DRS spectra using the Kubelka-Munk (KM) function, the Tauc plot, and the first derivative of the reflectance for a comparative study. $BiMnO_3$ exhibits an electronic transition with indirect band gap energy, $E_g \approx$ 0.92 eV. In contrast, all other functional perovskite oxides, namely bulk $V_2O_5$, $BaSnO_3$, $PbZr_{0.52}Ti_{0.48}O_3$, and $BiFeO_3$, show direct band gap interband transitions, with $E_g$ values of 2.27, 3.25, 3.10, and 2.48 eV, respectively. The estimated Urbach energy ($E_U$) values related to the induced defects in these direct band gap functional oxides are approximately 0.24, 0.38, 0.25, and 0.48 eV, respectively. Moreover, a reduction in the band gap energy of multiferroic $BiFeO_3$ was observed due to induced chemical pressure from (Ba, Ca)-doping and a decrease in particle size. Importantly, the evaluated band gap and Urbach energies of the functional perovskite oxide materials obtained from the analyses of the first derivative of reflectance and the Tauc plot method align remarkably well with the values deduced using the Kubelka-Munk function theory. Effectively, we propose a comprehensive electronic band structure for the multiferroic $BiFeO_3$, an important material for optoelectronic applications such as photovoltaic, photocatalytic, and photoferroelectric devices.

**TABLE OF CONTENTS**

    D. The DRS spectroscopy of pure and (Ba, Ca)-doped $BiFeO_3$ samples using the dilution method

    E. The optical spectroscopy of $BiFeO_3$ nanostructured sample

    F. The DRS spectroscopic studies on $BiFeO_3$ nanoparticle samples with different particle sizes

    G. Schematic band structure for the examined functional oxides and $BiFeO_3$

IV. CONCLUSION

## I. INTRODUCTION

In the $21^{st}$ century, there has been immense interest among researchers and industry experts in discovering new functional materials and engineering various materials to create synthetic functional materials.[1-3] This interest is driven by the rapid decline of our ecosystem as a consequence of our extensive reliance on conventional non-renewable energy sources.[1] Functional materials are generally defined as materials that perform specific functions in response to an applied stimulus.[3] Examples of functional materials include magnetic materials, ferroelectrics, multiferroics, piezoelectrics, thermoelectrics, semiconductors, ionic conductors, superconductors, carbon nanomaterials, and polymers.[4-6] The practical importance of these functional materials arises from their structural properties,[2] which involve the interaction of various degrees of freedom, such as spin, lattice, charge, and orbit, resulting in exceptional physical properties. Thus, functional materials are recognized as alternatives to conventional materials in areas such as energy storage, sensors, semiconductor technology, magnetocaloric applications, spintronics, and optoelectronic devices.[1-7]

Functional oxide materials with a perovskite structure (chemical formula $ABO_3$) represent a major group of materials that exhibit a variety of intriguing phenomena, including ferroelectricity, ferromagnetism, multiferroicity, colossal magnetoresistance (CMR), thermoelectricity, optoelectronics, and high-temperature superconductivity.[1-6] In this review article, we conducted a detailed investigation of the electronic energy levels and band structures of several selected perovskite oxide materials: vanadium pentoxide ($V_2O_5$), barium tin oxide ($BaSnO_3$), lead zirconium titanate ($PbZr_{0.52}Ti_{0.48}O_3$), bismuth manganite ($BiMnO_3$), and bismuth ferrite ($BiFeO_3$), using diffuse reflectance spectroscopy (DRS) in conjunction with the Kubelka-Munk theory. We specifically focused on examining the electronic structures of the multiferroic $BiFeO_3$ in both bulk and nanoscale forms due to its multifunctional characteristics, which include magnetoelectric, piezoelectric, and optoelectronic properties. It is essential to avoid the incorrect application of the Kubelka-Munk function and errors in estimating the band gap energy, as emphasized by Marotti *et al.*[8] and Makula *et al.*[9] in their recent reports. This study will also facilitate the probing of defects in the selected oxide materials, considering that optical spectroscopy is highly sensitive to disorders such as thermal, structural, and compositional changes within a material.[7] We assessed the defect states by computing the Urbach energy ($E_u$) for these materials, as disorder may impact their electronic structure, thereby affecting the functionality of these materials in various applications.

**A. Introduction to Functional Oxide Materials**

*1. Vanadium pentoxide:* Vanadium pentoxide ($V_2O_5$) is a multifunctional material with a layered orthorhombic structure, making it suitable for various practical applications such as electrochromics, thermoelectrics, sensors, solid-state batteries, and optoelectronic devices.[10-13] The optical properties including absorption, reflectance, and transmittance are significantly influenced by its structural features, as it can exist in different stable structures like orthorhombic ($α$-$V_2O_5$) and monoclinic ($β$-$V_2O_5$) forms.[14] The $α$-$V_2O_5$ material typically exhibits *n*-type semiconductor behavior with a band gap in the range of 2.2–2.6 eV,[14-17] primarily due to variations in its crystal structure, morphology, and doping elements.[14] The tunable optical properties of $α$-$V_2O_5$ make it an exceptional material for optoelectronic applications such as electrochromic, photochromic, and photocatalysis devices. In this study, we investigated the optical properties of sol-gel synthesized $α$-$V_2O_5$ powder using diffuse reflectance measurements. We found that the $V_2O_5$ powder heated at 450 °C is a single-phase $α$-$V_2O_5$ exhibiting an orthorhombic structure (space group *Pmn2₁*, see Fig. S1 in the supplementary material). A detailed analysis of the structural and electrical properties of the $α$-$V_2O_5$ powder will be presented in another publication.

***2. Barium tin oxide:*** Barium tin oxide ($BaSnO_3$) is a cubic perovskite wide band gap semiconductor material with a space group of $Pm\bar{3}m$.[18] It displays *n*-type semiconducting properties with a band gap between 3.1–3.4 eV.[18-20] Consequently, this material has found applications in various fields, such as dye-sensitized solar cells, thermally stable capacitors, and sensors.[19-21] Recently, Janifer *et al.*[22] detected two new intra-band transitions at 2.46 and 2.55 eV, which were ascribed to the transitions from the donor states associated with oxygen vacancies to the valence band of bulk $BaSnO_3$, exhibiting a band gap of 3.14 eV. Balamurugan *et al.*[19] investigated the optical properties of bulk $BaSnO_3$, which possesses a band gap of approximately 3.1 eV. Their sample was prepared using the solid-state reaction method, and characterized and reported elsewhere.[19] In this study, we have further analyzed the diffuse reflectance spectrum of bulk $BaSnO_3$ using the Kubelka-Munk function.

***3. Lead zirconium titanate:*** Lead zirconium titanate (PZT) is an excellent piezoelectric material that can be altered by modifying the Zr/Ti ratio.[23] Similarly, its band gap changes considerably from 3.2 eV to 3.7 eV with changes in the Zr/Ti ratio, as well as its crystalline structure and the calcination temperature.[23-26] This tunable optical property of PZT-based materials can be applied in optoelectronic fields such as optical displays, electro-optical devices, and photo-integration devices. In this study, we explored the optical properties of the compound $PbZr_{0.52}Ti_{0.48}O_3$ (PZT52/48) close to the morphotropic phase boundary (MPB) and sintered at 1200 °C for 12 hours. The X-ray diffraction pattern for the inspected PZT52/48 ferroelectric ceramic is presented in the supplementary material (refer to Fig. S2), showing the presence of both tetragonal and rhombohedral phases in this MPB compound as expected.[24]

***4. Bismuth manganite:*** Bismuth manganite ($BiMnO_3$) is a robust ferromagnetic insulator among the transition-metal perovskite oxides possessing a Curie temperature of 105 K.[27] The band gap values reported for this room-temperature ferroelectric $BiMnO_3$ range between 0.75 and 1.25 eV.[27-31] When $BiMnO_3$ is combined with $BiFeO_3$, the band gap drastically reduces from 2.7 eV ($BiFeO_3$) to 1.1 eV ($BiMnO_3$), creating possibilities for photoferroelectric devices.[29] The bulk $BiMnO_3$ material is analyzed here using diffuse reflectance spectroscopy to assess its optical property alongside the electronic structure of multiferroic $BiFeO_3$ and other functional oxides inspected, as it has an indirect band gap,[31] differing from the materials studied in the work.

**5. Bismuth ferrite:** Bismuth ferrite ($BiFeO_3$) is considered a model multiferroic compound because it displays spontaneous polarization and magnetization at room temperature and above.[32-35] In addition to magnetoelectric investigations,[34-37] this substance has also been explored for novel optoelectronic applications such as the photovoltaic effect,[38] photocatalytic effect,[39] optical second harmonic generation,[40] and generation of terahertz radiation.[41] In this context, Ramachandran *et al.*[42,43] investigated the optical characteristics of pure and (Ca, Ba)-doped $BiFeO_3$ ceramic materials using diffuse reflectance spectroscopy (DRS) employing the Kubelka-Munk theory. Ramachandran *et al.*,[42] were the pioneers in utilizing diffuse reflectance spectroscopy with a dilution technique for functional materials that eliminates the specular reflection of the incident photons.[42-44] This dilution method enabled them to analyze the DRS spectrum using the Kubelka-Munk function concerning the wavelength or energy of the incident photons.[44]

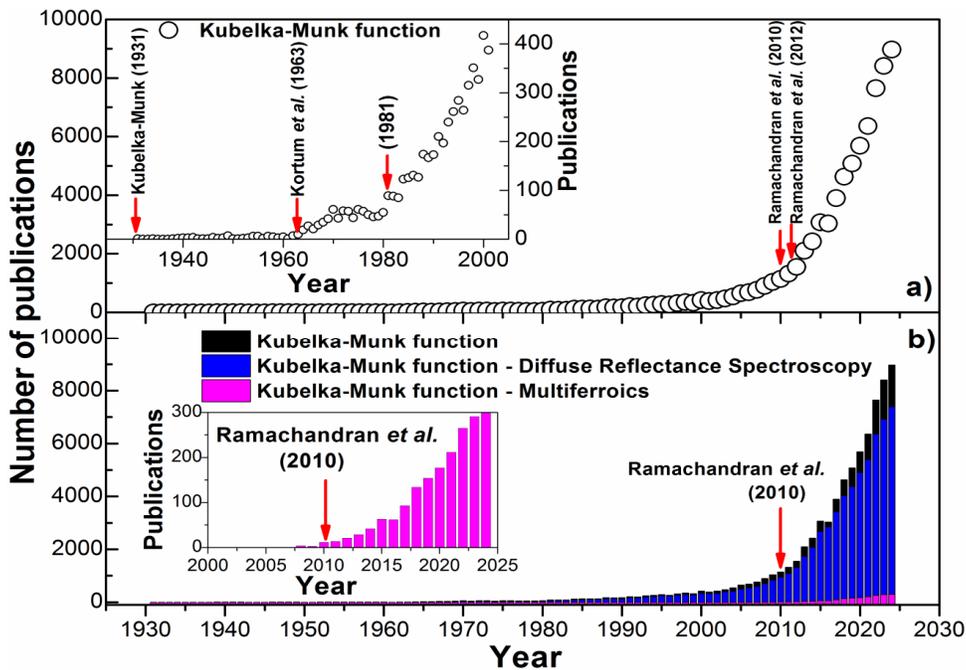

Figure 1 a) A survey of the number of publications on the topic ″Kubelka-Munk function″ and b) the number of publications versus the year of publication based on the surveys on the topics ″Kubelka-Munk function″, ″Kubelka-Munk function with Diffuse Reflectance Spectroscopy″, and ″Kubelka-Munk function with Multiferroics″ for each year since 1931, obtained via an estimation using the search engine Google Scholar. The insets of Figs. 1a and 1b provide a closer look at the surveys from the years 1931 to 2000 and from the years 2000 to 2024, respectively.

Notably, the study carried out by Ramachandran *et al.*[42] regarding charge transfer and electronic transitions in the multiferroic material BiFeO$_3$ is highlighted as one of the novel discoveries in the Scholarly Editions eBook focusing on "Issues in General Physics Research" (2011).[45] The effects of their contributions[42,43] are illustrated in Figure 1a, which presents a review of publication counts for the search term "Kubelka-Munk function" for each year since Kubelka and Munk published their original paper in 1931.[46] From this analysis, we noticed a slight increase in the number of publications (from 10 publications in 1963 to 61 publications in 1970, as depicted in the inset of Fig. 1a) after the work by Kortum *et al.*[47] in 1963. Afterward, there was another surge in annual publications starting in 1981, from about 100 publications in 1981 to several hundred by 2000. Finally, the steep rise in publications regarding the Kubelka-Munk function began in 2010; the publication count jumped from 1140 in 2010 to 9400 in 2024, which coincided with the reports of Ramachandran *et al.*[42,43] on the model multiferroic BiFeO$_3$ in the years 2010 and 2012. Comparable trends were also noted for the search topics of "Kubelka-Munk function with Diffuse Reflectance Spectroscopy" and "Kubelka-Munk function with Multiferroics", which are depicted in Fig. 1b. These findings demonstrate the importance of the research by Ramachandran and co-workers,[42,43] particularly the work referenced as 42.

**B. The Kubelka-Munk theory of diffuse reflectance for a semi-infinite layer**

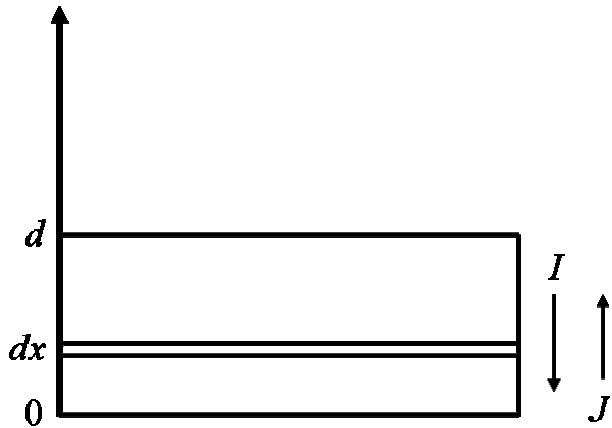

Figure 2 A parallel layer of thickness (*d*) of the sample used to derive the Kubelka-Munk formula. Reproduced from Reference 44.

We have reproduced the derivation of the Kubelka-Munk function for a semi-infinite layer to assist readers.[44] Consider a plane of the parallel layer with a thickness of *d*, capable of both scattering and absorbing radiation, being illuminated in the negative *x*-direction with a diffuse, monochromatic radiation flux, $I_{(x=d)}$ (Fig. 2). Assuming that the extension of the layer in the *yz*-plane is considerably larger than *d* so that edge effects can be ignored. First, let us consider an infinitesimal layer of thickness, *dx* parallel to the surface. The radiation flux in the negative x-direction is denoted as *I* and that in the positive x-direction, due to scattering is denoted as *J*. Here, the change in intensity of the incident radiation in the layer element *dx* is made up of three components as written below

$$-dI = -2kIdx - 2sIdx + 2sJdx \qquad (1)$$

Similarly, the decrease in intensity of the scattered radiation in the positive x-direction is

$$dJ = +2kJdx + 2sJdx - 2sIdx \qquad (2)$$

Where *k* represents the absorption coefficient of the material and *s* represents its scattering coefficient per unit length. By assuming these values $2k = K$ and $2s = S$, the two fundamental simultaneous differential equations that characterize the absorption and scattering processes can be derived as

$$\frac{dI}{dx} = -(K+S)I + SJ \qquad (3)$$

and

$$\frac{dJ}{dx} = (K+S)J - SI \qquad (4)$$

Further, the equations (3) and (4) can be written in the form using the term,

$$-\frac{dI}{Sdx} = -aI + J \qquad (5)$$

and

$$\frac{dJ}{Sdx} = -aJ + I \qquad (6)$$

Where the parameter, $a = \frac{S+K}{S} = 1 + \frac{K}{S}$. By dividing equation (5) by $I$ and equation (6) by $J$, and subsequently adding them with the assumed term $r = \frac{J}{I}$, we obtain

$$\frac{dr}{Sdx} = r^2 - 2ra + 1 \text{ or } \int \frac{dr}{r^2 - 2ra + 1} = S \int dx \qquad (7)$$

When integrating over the complete thickness of the layer, represented as $d$, two boundaries need to be taken into account: $x = 0$: $(J/I)_{I=0} = R_g$ = reflectance of the background and $x = d$: $(J/I)_{I=d} = R$ = reflectance of the sample. Using partial fractions, integrating equation (7) yields the relation:

$$\ln \frac{(R-a-\sqrt{a^2-1})(R_g-a+\sqrt{a^2-1})}{(R_g-a-\sqrt{a^2-1})(R-a+\sqrt{a^2-1})} = 2Sd\sqrt{a^2-1} \qquad (8)$$

When $d = \infty$, that is, when the layer possesses infinite thickness, $R_g = 0$, and solving for $R_\infty$ produces the following relation:

$$R_\infty = \frac{1}{a+\sqrt{a^2-1}} = a - \sqrt{a^2-1} = \frac{S}{S+K+\sqrt{K(K+2S)}}. \qquad (9)$$

Where, $R_g$ denotes the diffuse reflectance of the sample that can be easily measured and is solely dependent on K/S, as it is influenced exclusively by the ratio of the reflectance of an ideal white material to the absorbing impurities (samples being tested). For this reason, the value of $R_g = 1$ is nearly never achieved. If we solve the K/S relation, the equivalent solution is provided by

$$\frac{K}{S} = \frac{(1-R_\infty)^2}{2R_\infty} = F(R_\infty) \qquad (10)$$

or another equivalent formula was derived without the specular (mirror-like) reflection from the top surface of the sample by Myrick et al.[48] as

$$\frac{K}{S} = \frac{(1-R_\infty)^2}{4R_\infty} = F(R_\infty) \qquad (11)$$

The equation (10) is known as the Kubelka-Munk (KM) function.[46-48] This equation defines the reflectance of a semi-infinite sample with negligible specular reflection. It has been shown that the K and S values depend on the intrinsic absorption coefficient ($\alpha$) and intrinsic scattering coefficient ($s$) of the material, as revealed in a study by Yang *et al.*[49] The $F(R_\infty)$ spectrum is analogous to absorbance measurement in transmission spectroscopy, as explained by Milosevic and Berets.[50] They elucidated the concept using a scattering coefficient model, indicating that the scattering coefficient ($s$) of the material corresponds to the ratio of reflectance ($R$) and the grain size of the material ($d$), i.e., $s = \frac{R_\infty}{d}$, when the grain sizes of the sample are significantly small compared to the wavelength of incident photons. In this framework, they posited that the grain size ($d$) and the inter-grain distance within the material are identical since powder (polycrystalline ceramic) samples usually consist of randomly shaped grains. By applying the relationships: $s = \frac{R_\infty}{d}$ and $\alpha \approx kd$, equation (10) can be restructured as

$$\frac{k}{s} = \frac{kd}{R_\infty} = F(R_\infty) \qquad \text{or} \qquad \alpha \approx F(R_\infty) \times R_\infty \qquad (12)$$

From this stage onwards, we will use equation (12) to determine the band gap energy of the material using the measured diffuse reflectance data. In the following results and discussion, we will use the symbol $R$ instead of $R_\infty$ for the sake of convenience.

Diffuse reflectance ($R$) indicates the angular distribution of reflected photons, remaining unaffected by the angle of incidence.[44,51] Diffuse reflectance spectroscopy (DRS) is commonly employed for both quantitative and qualitative assessments of powders and uneven surface solids. It is an excellent sampling method for powdered or crystalline materials within the UV-visible-NIR spectral range. However, DRS spectroscopy has been measured using two slightly different approaches: *true* diffuse reflectance and *in-line* diffuse reflectance.[44] True diffuse reflectance measurement optimizes the collection of scattered photons while reducing the collection of specular (normal) reflected photons, thereby adhering closely to the Kubelka-Munk theoretical requirements. Conversely, in-line diffuse reflectance measurement captures both scattered and specular photons. Although it does not strictly qualify as diffuse reflectance, this technique is often used for swift spectral acquisition of a material. In this instance, we chose the former method for the reasons stated above.

## C. Determination of band gap using the spectrum of absorption coefficient

We evaluated the absorption coefficient ($\alpha$) as it varies with wavelength ($\lambda$) or energy ($E=h\nu$) by utilizing equation (12). Following this, we generated graphs of $(\alpha E)^2$ and $(\alpha E)^{1/2}$ versus incident photon energy ($E$) for all samples being investigated. The purpose of these plots is to determine if the $\alpha E$ data from a specific sample adheres to the formula:[52-55]

$$\alpha h\nu = A(h\nu - E_g)^n \text{ or } (\alpha E)^{1/n} = B(E - E_g) \quad (13)$$

In this equation, $A$ and $B$ denote proportionality constants, $E$ signifies the photon energy, $E_g$ refers to the band gap energy of an allowed transition, and $n$ is an exponent that indicates whether the band gap transition is direct ($n = 1/2$) or indirect ($n = 2$). We have employed both the Tauc plot with a baseline approach[9] and the first derivative reflectance ($\frac{dR}{d\lambda}$) technique[8] to precisely determine the band gap energy of the studied functional perovskite oxides and also to compare the findings with their estimated $E_g$ using the KM function.

## D. Determination of Urbach energy

The exponential behavior near the absorption edge of crystalline and amorphous materials, known as the Urbach tail band, is caused by structural disorders such as defects and grain boundaries in the sample.[53] This can be quantified using the Urbach empirical formula[52-55] as follows:

$$\alpha = \alpha_0 e^{\frac{E}{E_U}} \quad (14)$$

Here, $E_U$ represents the Urbach energy and $α_0$ is a constant. The $E_U$ energy is linked to electron transitions to the Urbach tail state above the valence band from one of the conduction band states and/or from the Urbach tail state below the conduction band to one of the valence band states (as illustrated in Fig. 14). The $E_U$ value of the probed samples will be determined by plotting and linear fitting the exponential behavior of the graph $lnα$ versus $E$.[52-55]

## II.  EXPERIMENTS

The Ocean Optics Spectrometer (USB2000, USA) was used to measure the diffuse reflectance spectra, $R(λ)$, of all functional oxide materials within the UV-visible-NIR range of 200–900 nm. This spectrometer offers an optical resolution that varies from approximately 0.3–10 nm based on the grating and the size of the entrance aperture. The diffuse reflectance, $R$ of a particular material at room temperature was measured by placing a reflection probe at a 45° angle to the sample's surface to minimize the detection of normal reflections captured by the spectrometer. Mechanically polished pellets of bulk samples were mainly used for the $R(λ)$ measurements, while powder and nano samples were packed into disc shapes without any heating at high temperatures. All samples were shaped into discs with a diameter of about 10–12 mm and a thickness of approximately 1–2 mm, to comply with the requirement of using the KM function for analyzing the recorded $R(λ)$ spectra of the samples.[44] The samples were pelletized into disc shapes with a pressure of nearly 80 kg/cm$^3$ using a hydraulic press.[42,43] The diffuse reflectance of all functional oxides under study was recorded relative to the reflectance of an inactive, non-absorbing standard material, barium sulfate (BaSO$_4$). That is,

$$R' = \frac{R_{sample}}{R_{standard}} \qquad (15)$$

The absolute reflectance of the standard BaSO$_4$ was calibrated to unity across the whole wavelength range under study, thus the measured relative reflectance ($R'$) of the samples is effectively equivalent to their absolute reflectance ($R$).[44]

In our earlier studies,[42,43] we performed research on the diffuse reflectance spectra using the KM function, $F(R)$ of pure and (Ca, Ba)-doped BiFeO$_3$ polycrystalline samples. These investigations involved recording the $R(λ)$ spectra of the BiFeO$_3$ samples using a dilution method.[44,56] This approach was employed to eliminate specular reflection, facilitating the measurement of the absolute diffuse reflectance of the samples. In this process, 5–10 wt% of the test material was thoroughly blended with a non-absorbing standard material, barium sulfate ($BaSO_4$) by grinding. The resulting mixture was then formed into discs with a diameter of 10–12 mm and a thickness of 1–2 mm using a hydraulic press. Finally, the diffuse reflectance of these samples was recorded with respect to the pure standard.[44,47] By eliminating normal reflection using this method, we successfully analyzed the measured $R(\lambda)$ spectra of the samples based on the KM theory. As a result, in our prior works,[42,43] two charge transfer transitions and two doublet *d-d* transitions were experimentally detected for the pure and the doped $BiFeO_3$ polycrystalline samples.

In this study, we conducted further analysis of the $R(\lambda)$ data from the pure and doped $BiFeO_3$ samples using the dilution method by evaluating the first derivative of their reflectance, $\frac{dR}{d\lambda}(\lambda)$ to compare with the results derived from the relative reflectance data of the pristine sample. This comparison will highlight the importance of correctly employing the KM function to prevent miscalculation of the energy of the band gap and other electronic transitions of a testing material, as proposed by Marotti *et al.*[8] and Makula *et al.*[9] Notably, employing the first derivative of the reflectance analysis enabled us to remove the background absorbance in the relative reflectance measurement. Therefore, the suggested comparison of the energy of the electronic transitions obtained from the relative reflectance and dilution methods for the $BiFeO_3$-based materials provided a clear understanding of their electronic structure, as the experimental data is free from the effects of the background absorbance and specular reflection in the relative reflectance and dilution measurements, respectively. It should be acknowledged that the estimated relative error in the *F(R)* function, $\frac{dF(R)}{F(R)}$ for the functional oxides was about 1–2% within the range of 0.2 < R < 0.7 (see Fig. S3 in the supplementary material). This relative error significantly increases at lower or higher reflectance values, which is in agreement with the literature.[44]

## III. RESULTS AND DISCUSSION

### A. Optical properties of semiconductor compounds $V_2O_5$ and $BaSnO_3$:

Figure 3a displays the measured $R(\lambda)$ spectra of the prepared orthorhombic $\alpha$-$V_2O_5$ and commercial $V_2O_5$ powder materials using the method described in equation (15). The *R* values of the $\alpha$-$V_2O_5$ material are in the range of 0.1 < R < 0.7, while the *R* values of the commercial samples are in the range of 0.1 < R < 0.5. The optical absorption onset for both samples, corresponding to the band gap energy, is visible near $\lambda = 535$ nm. To accurately determine the optical band gap energy of both samples, we obtained the first derivative of the reflectance with respect to wavelength, i.e., $\frac{dR}{d\lambda}$.[8] It is expected that the $\frac{dR}{d\lambda}$ diverges near the wavelength of the band gap energy, $\lambda_g = \frac{hc}{E_g}$.[8,57,58] However, the structural defects and disorders in the material will result in a peak feature near $\lambda_g$ instead of the divergence feature.[8,58] Hence, we plotted a graph of $\frac{dR}{d\lambda}$ versus $\lambda$ of the prepared $\alpha$-V$_2$O$_5$ powder as a representative plot in the inset of Fig. 3a. We found a peak feature around $\lambda_g = 531.2$ nm, indicative of the electronic transition being a direct transition in the prepared $\alpha$-V$_2$O$_5$.

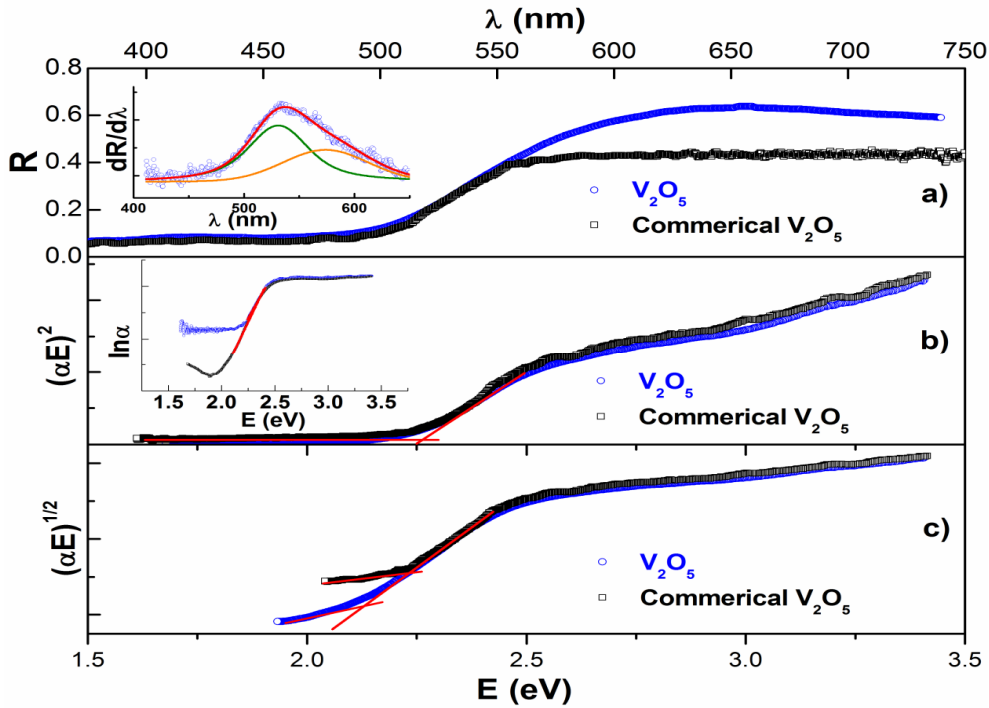

Figure 3 a) The diffuse reflectance ($R$) of the prepared $\alpha$-V$_2$O$_5$ and commercial V$_2$O$_5$ powder samples as a function of wavelength ($\lambda$), b) the plot of $(\alpha E)^2$ versus photon energy ($E$), and c) the plot of $(\alpha E)^{1/2}$ versus photon energy ($E$). The inset in Fig. 3a shows the graph of $\frac{dR}{d\lambda}$ versus $\lambda$ for the prepared $\alpha$-V$_2$O$_5$ and the inset in Fig. 3b displays the plot of $ln\alpha$ versus $E$ for both samples.

The commercial samples also show a peak feature near $\lambda_g = 529.0$ nm in the $\frac{dR}{d\lambda}(\lambda)$ data (see Fig. S4 in the supplementary material). After energy conversion of the wavelength of absorption edges, the optical direct band gap energy ($E_g$) is estimated to be nearly equal ($E_g \sim 2.34$ eV) for both prepared and commercial V$_2$O$_5$ materials. This direct transition is essentially attributed to a transition from the V 3*d* orbital-dominated conduction band to the valence band made of O 2*p* orbitals.[10] These observations are in good agreement with the literature.[14-17,59,60] Additionally, we noted another peak at 574.0 nm (~2.16 eV) for the prepared α-$V_2O_5$ sample, which could be due to transitions from donor levels induced by the defects.[8,58] This is analyzed further using the KM function in the following.

We calculated the KM function, *F(R)* as a function of energy (*E*) using equation (12) and subsequently determined the absorption coefficient spectrum, *α(E)*, for both the prepared and commercial $V_2O_5$ samples (see Fig. S5 in the supplementary material). To evaluate their band gap energy ($E_g$) using equation (13), we plotted the Tauc graphs of $(\alpha E)^2$ and $(\alpha E)^{1/2}$ versus *E* and presented them in Fig. 3b and 3c, respectively. The estimated energy of the optical band-edge was approximately 2.27 eV for both the α-$V_2O_5$ and commercial $V_2O_5$ samples using the best linear fit to the Tauc plots (see Fig. 3). This finding is consistent with the observation made using the absorption onset ($E_g$~2.34 eV) with the assistance of the $\frac{dR(\lambda)}{d\lambda}$ spectrum. In addition, we determined the indirect band gap energy ($E_g$) to be 2.13 and 2.23 eV for the α-$V_2O_5$ and commercial $V_2O_5$ samples, respectively. We attribute the indirect band gap transitions to induced structural defects such as oxygen vacancies and grain boundaries.[8,57,58]

To further investigate this finding, the exponential behavior near the absorption edge was examined using the Urbach empirical formula described in equation (14). A graph of *lnα* versus *E* is displayed in the inset of Fig. 3b. The Urbach energy ($E_U$ = 0.24 eV) for both systems was determined by calculating the inverse of the obtained slope using linear regression, depicted as the straight line in the inset of Fig. 3b. This value is relatively small compared to the value $E_U$ = 0.5 eV of the 253 nm thick $V_2O_5$ film.[61] Hence, we argue that the defect-induced Urbach tail states in our sample (as well as commercial $V_2O_5$) are shallow trap states when compared to the $V_2O_5$ film.[61] Notably, the energy required for the transitions from the conduction band minimum (CBM) to the Urbach band above the valence band and from the Urbach band below the conduction band to the valence band maximum (VBM) is about ~2.1 eV which is found to be nearly equal to the obtained indirect band gap energy (2.13 eV) of our α-$V_2O_5$ (refer to the band structure in Fig. 15a).

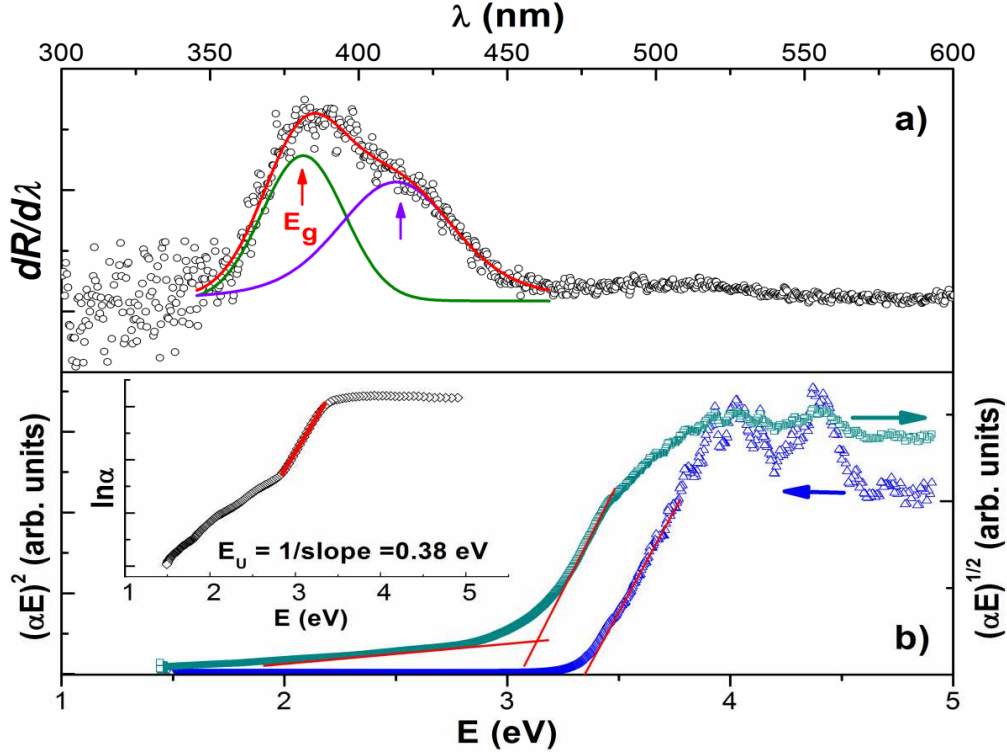

Figure 4 a) A plot of $\frac{dR}{d\lambda}$ against $\lambda$ for bulk $BaSnO_3$ sample and b) the plots of $(\alpha E)^2$ (left y-axis) and $(\alpha E)^{1/2}$ (right y-axis) versus $E$ for this sample. The inset shows a plot of $ln\alpha$ versus $E$ for the sample.

The recorded $R(\lambda)$ spectrum of the bulk $BaSnO_3$ sample presented in our earlier work[19] is shown in Fig. S6 in the supplementary material. It is observed that the diffuse reflectance ($R$) lies in the range of $0.1 < R < 0.8$ over the UV-Vis-NIR region with a sharp absorption edge near the band gap transition ($E_g = 3.1$ eV).[19] The estimated KM function against energy using equation (12) is plotted in the inset of Fig. S5. To check the accuracy of the estimated $E_g$ value of bulk $BaSnO_3$, we calculated the reflectance's first derivative, $\frac{dR(\lambda)}{d\lambda}$, which is illustrated in Fig. 4a. From the observed peak feature near $\lambda_g = 381.3$ nm, we evaluated the band gap energy of bulk $BaSnO_3$ to be $E_g = 3.25$ eV, consistent with the literature.[18-20,62] This finding indicates the direct allowed band-to-band transition from the CBM (largely due to *Sn 5s* orbitals with a small contribution from *O 2s* orbitals) to the VBM (dominated by *O 2p* orbitals).[20] We also observed another feature at a higher wavelength of 412.6 nm (corresponding to $E = 3.0$ eV), likely due to structural disorders in this sample which will be inspected using the Urbach tail band (see the inset of Fig. 4b).

We constructed the Tauc plots, $(\alpha E)^2$ and $(\alpha E)^{1/2}$ versus $E$, using equation (12) and displayed them in Fig. 4b. Through linear regression on these Tauc plots, we determined the direct and indirect band-edge energies to be approximately 3.35 and 3.13 eV, respectively. It was observed that the obtained direct band gap value ($E_g$ = 3.35 eV) of bulk BaSnO$_3$ aligns with reported values (3.1–3.4 eV),[18-20] and is comparable to the value ($E_g$ = 3.25 eV) obtained using the $\frac{dR}{d\lambda}(\lambda)$ spectrum in Fig. 4a. The difference in the direct and indirect transition onsets for bulk BaSnO$_3$ is about 0.22 eV, which is lower than those found in literature, experimental (~0.6 eV)[63,64] and theoretical (~0.5 eV) values.[64,65] To investigate the Urbach tail band in bulk BaSnO$_3$, we created a plot of $ln\alpha$ as a function of $E$, presented in the inset of Fig. 4b. By utilizing a linear fit of the exponential variation around the band gap absorption edge, depicted as a straight line in the inset of Fig. 4b, the Urbach energy ($E_U$) was determined to be approximately $E_U$ = 0.38 eV for bulk BaSnO$_3$. It is comparable to reported values ($E_U$ = 0.25–0.34 eV) of BaSnO$_3$-based systems.[66,67] We also estimated the separation energy of $E_C$-$E_U$ to be ~2.97 eV, which coincides with the lower energy peak feature at 3.0 eV in the $\frac{dR(\lambda)}{d\lambda}$ spectrum (see Fig.4a). These observations suggest that the estimated Urbach energy is due to the lowest energy level of the Urbach tail band formed in BaSnO$_3$ due to induced defects such as oxygen vacancies and grain boundaries, leading to the creation of localized energy states within the energy gap.[67]

**B. The optical spectroscopy study on piezoelectric bulk PZT(52/48)**

The evaluated $F(R)$ versus $E$ and $\frac{dR(\lambda)}{d\lambda}$ spectra for the bulk PZT(52/48) were displayed in Fig. S7 in the supplementary material. It is noted that the diffuse reflectance ($R$) of the sample is in the range of 0.1 < $R$ < 0.9 over the measured wavelength range, 200–800 nm. Two peak features were observed in the $\frac{dR(\lambda)}{d\lambda}$ spectrum with a first peak at $\lambda_g$ = 390.4 nm corresponding to the optical band gap absorption onset ($E_g$ ~3.18 eV). This direct electronic transition is primarily due to the electronic interband transition from the Pb 6$p$ states dominated CBM to the VBM due to the hybridization of Pb 6$s$ and O 2$p$ orbitals.[26] The second peak feature at 440.8 nm ($E$ = 2.81 eV) is likely due to induced defects, which is further investigated using the Urbach energy. The obtained Tauc plots (($\alpha E)^2$ and $(\alpha E)^{1/2}$ versus $E$) using equations (12) and (13) were presented in Fig. 5. By using the best linear fit to the Tauc plots near the absorption edge which is illustrated as the straight lines, we deduced the direct and indirect band gap energies to be about ~3.1 and

~2.7 eV, respectively, for bulk PZT(52/48). The obtained energy (3.1 eV) of the absorption edge is comparable to the value ($E_g$ = 3.18 eV) obtained from the $\frac{dR(\lambda)}{d\lambda}$ spectrum. This result is in good agreement with the reported values for the PZT(53/47) and PZT(52/48) materials.[24,68] We also determined the Urbach energy ($E_U$ = 0.25 eV) of the ceramic PZT(52/48) using the exponential behavior of the $ln\alpha(E)$ plot near the optical absorption edge which is presented in the inset of Fig. 5. This is essentially due to the localized energy states that form the Urbach tail band, commonly lying below the CBM and/or above the VBM.[69]

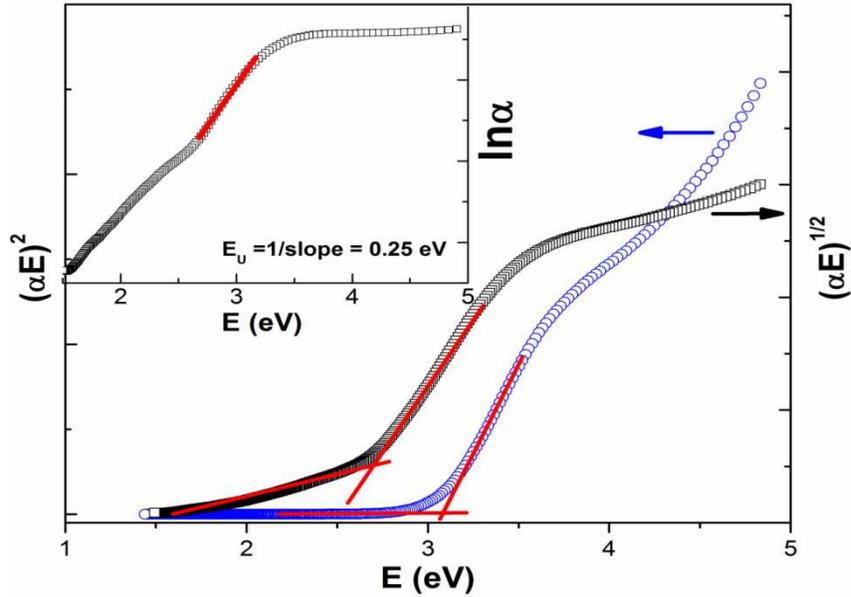

Figure 5 The Tauc plots of $(\alpha E)^2$ (left *y*-axis) and $(\alpha E)^{1/2}$ (right *y*-axis) versus *E* for bulk PZT(52/48). The inset shows a graph of *lnα* plotted against *E* for the sample.

### C. The DRS spectroscopy of multiferroic bulk BiFeO$_3$ and BiMnO$_3$

Figure 6 illustrates the recorded $R(\lambda)$ data of bulk BiFeO$_3$ and BiMnO$_3$ samples using relative reflectance measurement as described in equation (15). The measured *R* value of BiFeO$_3$ is in the range from 0 to 0.6, while that of BiMnO$_3$ is in the range from 0 to 0.2. To estimate the band gap energy of these samples, we evaluated the first derivative of reflectance against wavelength, $\frac{dR(\lambda)}{d\lambda}$, shown in the inset of Figs. 6a and 6b. A peak at a lower wavelength of 365.3 nm (~3.39 eV) corresponds to the charge transfer excitation to the Bi 6*p* states from the O 2*p* states. This value is comparable to the value (3.6 eV) estimated using the *F(R)* spectrum recorded with the dilution method for the same sample.[42] Additionally, four peaks related to two doubly degenerate *d-d* transitions are observed at ~2.32, ~2.16, ~1.75, and ~1.66 eV (corresponding to 535.2, 574.3, 706.6, and 748.0 nm), in line with the results of the dilution method detailed in our earlier work.[42] However, there is no peak feature in the $\frac{dR(\lambda)}{d\lambda}$ spectrum of BiMnO$_3$ (the inset in Fig. 6b), confirming that this compound has an indirect electronic transition.[27-31]

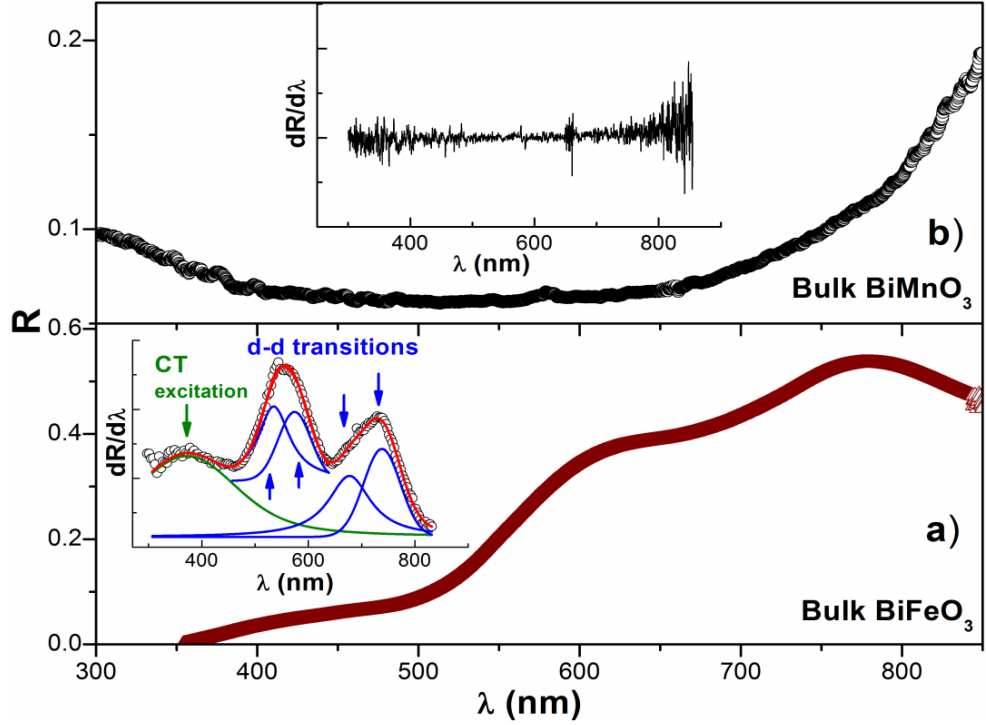

Figure 6 a) The *R(λ)* spectra of a) BiFeO$_3$ and b) BiMnO$_3$ bulk materials in the wavelength range of 350–850 nm. The inset in Fig. 6a shows the deconvoluted $\frac{dR(\lambda)}{d\lambda}$ spectrum of BiFeO$_3$ and the inset in Fig. 6b displays the data of $\frac{dR}{d\lambda}$ versus λ of BiMnO$_3$.

To thoroughly analyze these outcomes, we estimated the factors $(\alpha E)^2$ and $(\alpha E)^{1/2}$ versus *E* for both BiFeO$_3$ and BiMnO$_3$ materials (Fig. 7), using their *F(R)* spectra illustrated in Fig. S8 in the supplementary material. By using linear regression to plot $(\alpha E)^2$ versus *E* around the electronic transitions, a feature at a higher energy of 3.1 eV due to the allowed *p-d* charge transfer band-band transition with an optical absorption edge at 2.40 eV was observed for bulk BiFeO$_3$. Another feature at 2.18 eV is also seen, which is consistent with the energy (~2.22 eV) related to one of the four *d-d* transitions for BiFeO$_3$ obtained using its $\frac{dR(\lambda)}{d\lambda}$ spectrum (see the inset of Fig. 6a). These observations are similar to our earlier DRS study using the dilution method.[42] Meanwhile, the indirect transition in $BiFeO_3$ was detected at approximately 1.86 eV, consistent with the literature value (1.84 eV).[40] For $BiMnO_3$, no feature related to the direct band gap transition (Fig. 7b) was detected as expected,[31] which is also evident from our $\frac{dR(\lambda)}{d\lambda}$ data (see the inset of Fig. 6b). However, the indirect band gap transition at ~0.92 eV found in the curve of $(\alpha E)^{1/2}$ versus $E$ is observed for $BiMnO_3$, in line with the literature.[27-31]

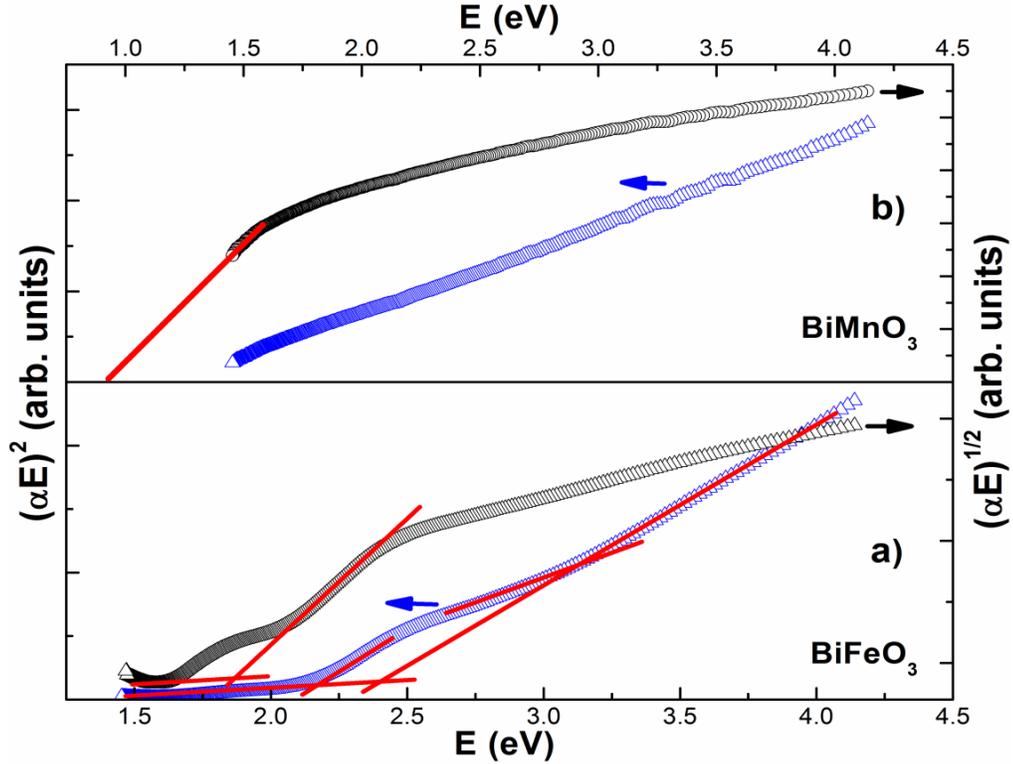

Figure 7 The graphs of a) $(\alpha E)^2$ (left y-axis) and b) $(\alpha E)^{1/2}$ (right y-axis) versus $E$ for the $BiFeO_3$ and $BiMnO_3$ samples.

To determine the $E_U$ values for both compounds $BiMnO_3$ and $BiFeO_3$, the exponential behaviors near their absorption edges were fitted using linear regression, as shown by the straight lines in Fig. 8. The evaluated $E_U$ value is about 0.67 eV for $BiMnO_3$. For $BiFeO_3$, the obtained $E_U$ value is approximately 0.48 eV, consistent with literature values (0.5–0.7 eV).[70-73] We attribute the observed energy of the Urbach tail band in $BiFeO_3$ to the moderate trap level created by oxygen vacancies that lie in the energy range of 0.4–0.9 eV.[72,73] All the observed electronic transitions in bulk BiFeO$_3$ found using the relative diffuse reflectance described in equation (15) are compared with the results obtained using the dilution method described in section D.

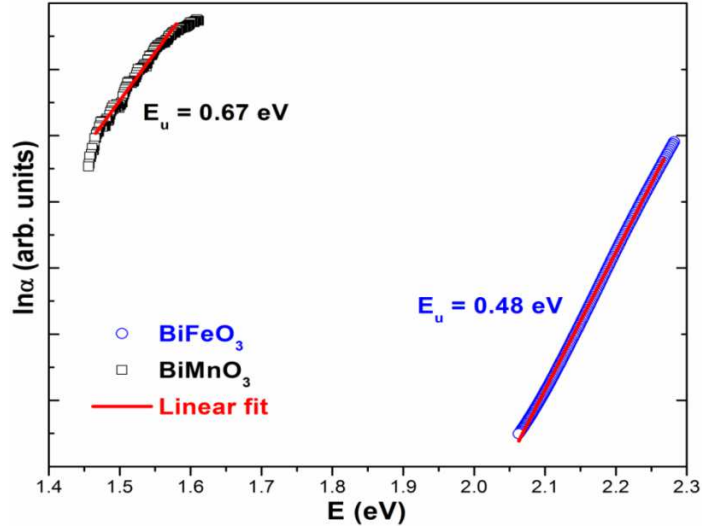

Figure 8 The exponential behaviors (*lnα* versus *E*) of the Urbach tail bands in the BiFeO$_3$ and BiMnO$_3$.

## D. The DRS spectroscopy of pure and (Ba, Ca)-doped BiFeO$_3$ samples using the dilution method

Figure 9a displays the graphs of $\frac{dR}{d\lambda}$ versus *E* for pure and (Ba, Ca)-doped BiFeO$_3$ ceramic samples obtained using the *R(λ)* data recorded by the dilution method reported in our earlier works.[42,43] This analysis was performed to reaffirm our previous findings and compare the outcomes with the results obtained using the measured relative reflectance data of pure BiFeO$_3$ (see Figs. 6-8). Features related to direct band gap transition and two doubly degenerate *d-d* ($^6A_{1g} \rightarrow {}^4T_{2g}$) transitions were observed for all BiFeO$_3$ samples. Additionally, features related to charge transfer excitations are seen for these materials in the energy range of 3.0–5.0 eV. To determine the energies of all the observed electronic transitions, we applied a deconvolution method to the $\frac{dR}{d\lambda}(E)$ data of the BiFeO$_3$-based materials. As a representative plot, the deconvoluted $\frac{dR}{d\lambda}$ spectrum in the range of 1.5–3.0 eV of pure BiFeO$_3$ is presented in Fig. 9b. Notably, we observed a direct band gap transition at about 2.48 eV for the pristine sample, similar to the value (2.50 eV) obtained using its *F(R)* data.[42,43] Additionally, four *d-d* transitions allowed by spin-orbit coupling are detected at ~2.14, ~2.01, ~1.75, and ~1.64 eV in the pristine compound and these transitions are also observed in all the doped $BiFeO_3$ samples, consistent with the results obtained in their *F(R)* data.[42] Besides, the features of one *p-d* and two *p-p* charge transfer transitions are detected in the energy range of 3.0–5.0 eV. Particularly, the feature related to charge transfer excitation is observed at ~4.30 and ≥ 4.50 eV for pure and doped $BiFeO_3$ samples, respectively (see Fig. S9 in the supplementary material). These observations validate our findings using the KM function in our earlier studies.[42,43]

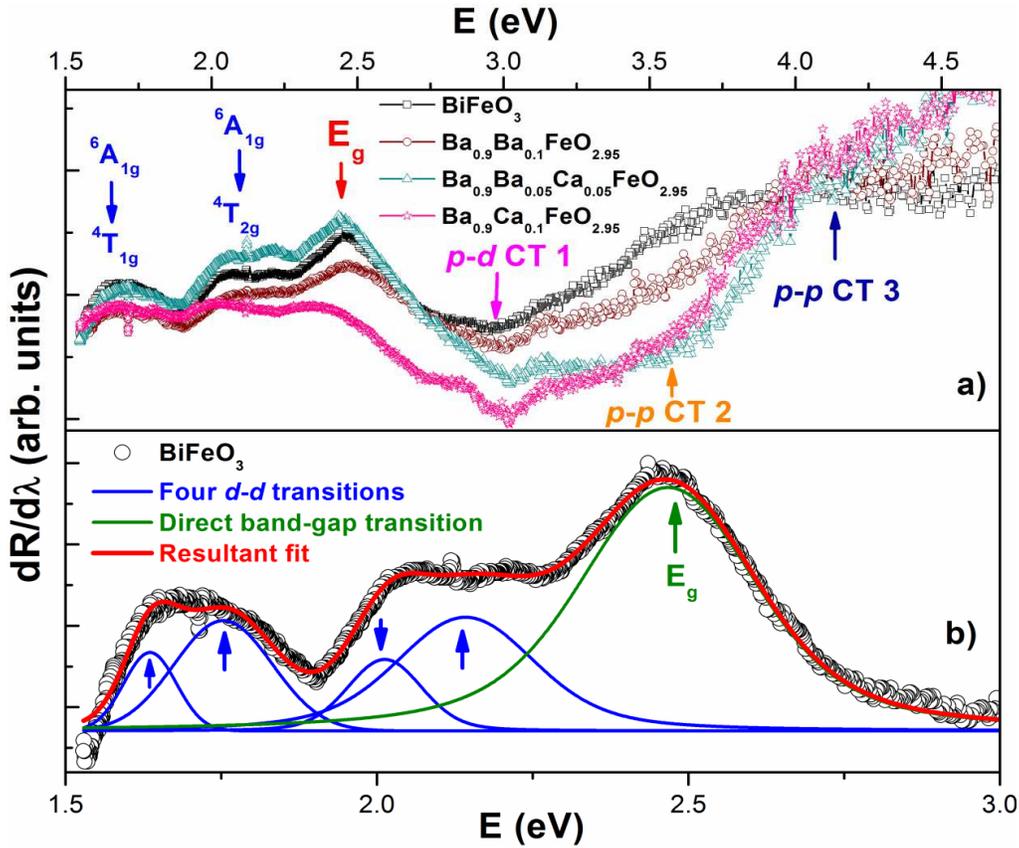

Figure 9 a) the $\frac{dR}{d\lambda}(E)$ spectra of pure and (Ba, Ca)-doped $BiFeO_3$ ceramic samples obtained using the dilution technique and b) the deconvoluted $\frac{dR}{d\lambda}(E)$ spectrum of pure $BiFeO_3$ as a representative graph.

The evaluated direct band gap transitions of the pristine and doped $BiFeO_3$ samples are consistent with our earlier findings,[42,43] and are listed in Table 1 along with their room-temperature dielectric constant ($\varepsilon_r$) at the selected frequencies of 30 kHz, 100 kHz, and 1 MHz presented in our low-temperature dielectric and magnetodielectric studies on these samples.[36,37,74] Interestingly, we noticed that the dielectric constant of doped BiFeO$_3$ samples increases with a decrease in their band gap energy (i.e., $E_g = \frac{1}{\varepsilon_r^2}$) when compared with pristine BiFeO$_3$, consistent with the recent observation in the Hf-substituted BaTiO$_3$ low-*k* systems according to a Bohr-like model.[75] This finding is essentially due to the dopant-induced defects forming trap-level states below the CBM in these examined BiFeO$_3$-based materials. Thus, the magnetic, dielectric, magnetoelectric, and magnon-phonon coupling were improved in the Ca-doped BiFeO$_3$ materials compared to the pristine compound.[36,37,76-78]

Table 1 the direct band gap energy and dielectric constant measured at different frequencies for pristine and doped BiFeO$_3$ ceramic samples.

| Sample | $E_g$ (eV) | $\varepsilon_r$ at 30 kHz | $\varepsilon_r$ at 100 kHz | $\varepsilon_r$ at 1 MHz |
|---|---|---|---|---|
| BiFeO$_3$ | 2.48 | 86.5 | 85.4 | 59.9 |
| Bi$_{0.9}$Ba$_{0.1}$FeO$_{2.95}$ | 2.46 | 232.9 | 230.4 | 170.0 |
| Bi$_{0.9}$Ba$_{0.05}$Ca$_{0.05}$FeO$_{2.95}$ | 2.43 | 566.7 | 469.6 | 385.9 |
| Bi$_{0.9}$Ca$_{0.1}$FeO$_{2.95}$ | 2.33 | 444.3 | 267.0 | 84.4 |

**E. The optical spectroscopy of BiFeO$_3$ nanostructured sample**

We synthesized self-assembled 3D flower-like nanostructures consisting of several nanorods with a length of about 1 μm and a thickness of about 100-200 nm (see the inset of Fig. 10a) of BiFeO$_3$ material using the ethylene glycol polyol process, as described by Zhong *et al.*[79] X-ray diffraction data of the BiFeO$_3$ nanostructured sample heated at 450 °C for 1 hour showed the presence of the secondary phase Bi$_2$O$_3$ along with the desired phase *R*3*c* BiFeO$_3$, as illustrated in Fig. S10 in the supplementary material. Detailed analysis of structural, dielectric, and ferroelectric properties will be published elsewhere. In this article, we analyzed its optical properties to compare them with the characteristics of bulk BiFeO$_3$. The measured *R(λ)* spectrum (left *y*-axis) and the obtained *F(R)* data (right *y*-axis) of the nanostructured BiFeO$_3$ are shown in Fig. 9a. The absolute *R* values of nanostructured BiFeO$_3$ are in the range of 0 to 0.7. To compare its characteristics with bulk BiFeO$_3$, we evaluated $\frac{dR}{d\lambda}$ as a function of wavelength (*λ*) and presented in Fig. 10b. Features related to direct band gap transition and four *d-d* transitions were observed at approximately 2.61, 2.20, 2.0, 1.86, and 1.69 eV, respectively. These findings are in excellent agreement with the data of bulk BiFeO$_3$ obtained using both relative reflectance and dilution measurements (see Figs. 6-8).[42] Our results are also consistent with a very recent report on the electroluminescence in BiFeO$_3$-based heterostructures by Simon *et al.*[80]

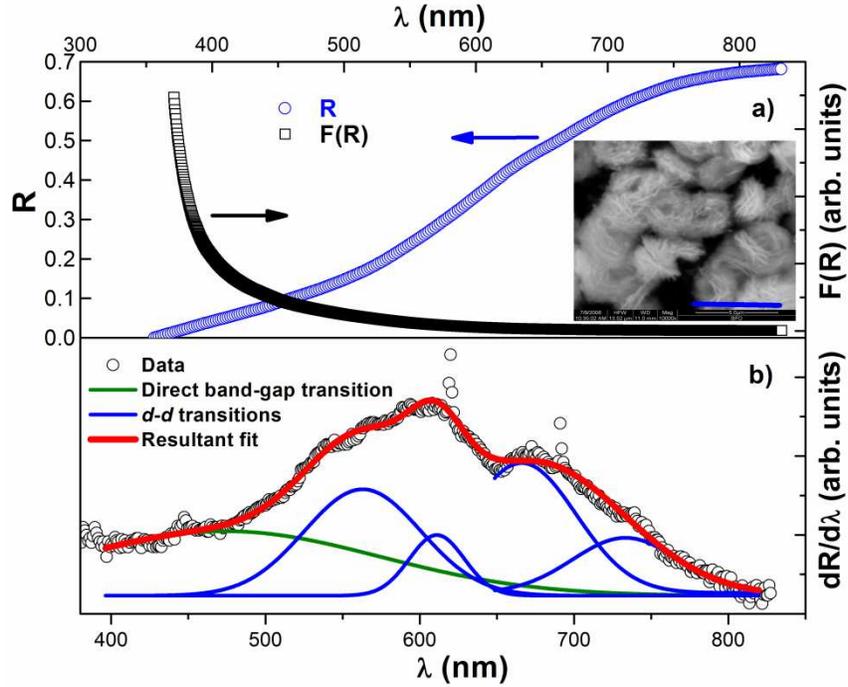

Figure 10 a) The reflectance spectrum, *R(λ)* (left *y*-axis) and *F(R)* spectrum (right *y*-axis) of the nanostructured BiFeO$_3$ and b) the deconvoluted $\frac{dR}{d\lambda}(\lambda)$ plot of the nanostructured sample showing possible transitions.

The Tauc plots for nanostructured BiFeO$_3$ made using equations (12) and (13) are shown in Fig. 11. We note that there is no clear feature related to the direct band gap edge in the graph of $(\alpha E)^2$ versus *E* (left *y*-axis), which could be due to induced defects including the secondary phase Bi$_2$O$_3$ which has a band gap in the wide range of 2.1–3.0 eV.[81] Thus, the band gap transition is found to be weaker in its $\frac{dR(\lambda)}{d\lambda}$ data (Fig. 10b) when compared to bulk BiFeO$_3$ (Fig. 8b). So, we roughly estimate the optical absorption edge to be at 2.4 eV, similar to bulk BiFeO$_3$ (see subsection C). Besides, the transitions related to a *d-d* transition and indirect band gap transition are detected at 2.0 and 1.69 eV, respectively. We drew the curve of *lnα* versus *E* to calculate the $E_U$ energy related to induced defects in the nanostructured BiFeO$_3$, as illustrated in the inset of Fig. 10. Our estimation gives the $E_U$ value of about 0.35 eV for this sample, which is slightly lower (0.48 eV) than that of bulk BiFeO$_3$ (see section 3.4). Notably, this value is comparable to literature values (0.4–0.9 eV) of BiFeO$_3$-based materials.[70-73]

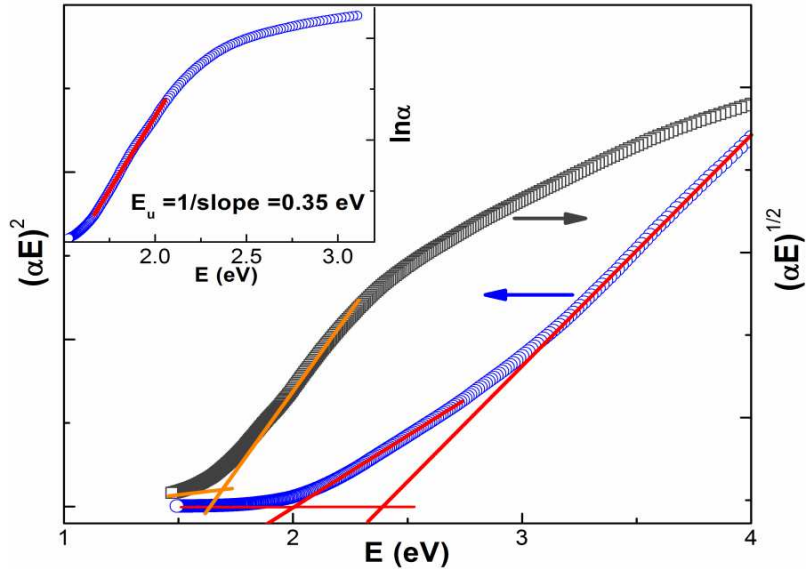

Figure 11 The Tauc plots of $(\alpha E)^2$ (left y-axis) and $(\alpha E)^{1/2}$ (right y-axis) versus E for the BiFeO$_3$ nanostructured sample. The inset displays the $ln\alpha$ versus E of this sample and the red solid line represents fitting to the Urbach tail band.

## F. The DRS spectroscopic studies on BiFeO$_3$ nanoparticle samples with different particle sizes

Figure 12 displays the first derivative spectra of reflectance, $\frac{dR}{d\lambda}(\lambda)$, of BiFeO$_3$ nanoparticle samples heated at various temperatures, 350–600 °C. These spectra were obtained using the $R(\lambda)$ spectra presented in a recent article by Sahoo and Dixit.[82] Their $R(\lambda)$ spectra were recorded using a Cary 4000 spectrometer with BaSO$_4$ as a standard. It is observed that spike-like divergences near the wavelength ($\lambda_g$) corresponding to the band gap energy ($E_g$) in the $\frac{dR(\lambda)}{d\lambda}$ data for the nanoparticle samples heated at high temperatures, specifically 500–600 °C, while the divergence of $\frac{dR}{d\lambda}$ is less pronounced for samples calcined below 500 °C (see inset of Fig. 12). Importantly, the $\lambda_g$ of the band gap edge of these samples was found to shift to a lower wavelength with increasing calcination temperature. To estimate the $\lambda_g$ value of these samples, we employed a deconvolution technique on the $\frac{dR(\lambda)}{d\lambda}$ plots of the BiFeO$_3$ nanoparticle samples presented in Fig. S11 of the supplementary material. The evaluated $E_g$ values of these samples vary from 2.01 eV to 2.22 eV with increasing average particle size (as heating temperature increases), as illustrated in Fig. 14a. Notably, these obtained $E_g$ values of BiFeO$_3$ nanoparticles were considerably red-shifted in comparison with its bulk counterpart ($E_g$ = 2.48 eV, see Table 1).[42,43]

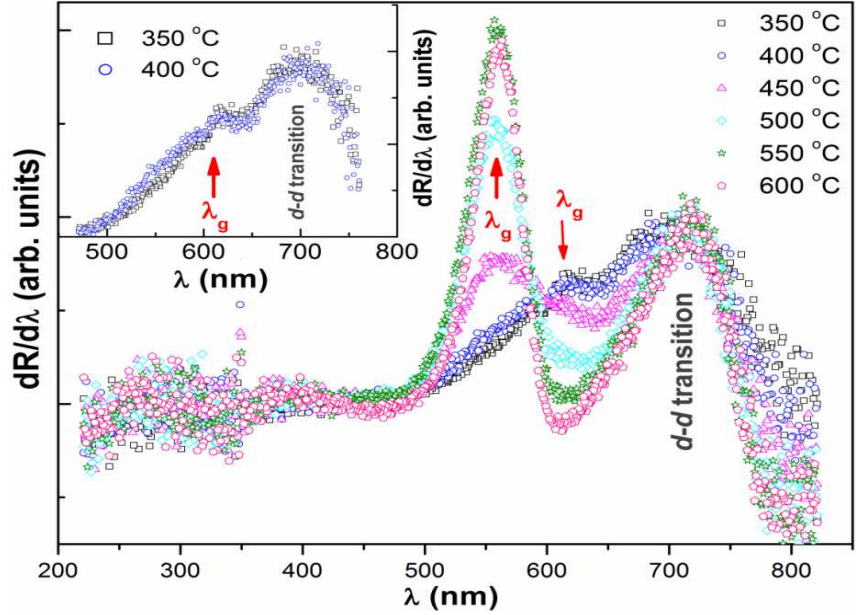

Figure 12 The curves of $\frac{dR}{d\lambda}$ versus $\lambda$ for the BiFeO$_3$ nanoparticles calcined at temperatures ranging from 350 to 600 °C. The inset displays the $\frac{dR(\lambda)}{d\lambda}$ data for the BiFeO$_3$ nanoparticles heated at 350 and 400 °C in an enlarged view.

The reduction of the $E_g$ value with decreasing size of BiFeO$_3$ nanoparticles was also observed in the constructed Tauc plots of $(\alpha E)^2$ and $(\alpha E)^{1/2}$ versus $E$ (Fig. 13). Particularly, the $E_g$ value is about 2.15 eV for the BiFeO$_3$ nanoparticle sample heated at 600 °C which tends to decrease slightly with decreasing the calcination temperature, T ≤ 550 °C. This finding is analogous to the results obtained from their spectrum $\frac{dR(\lambda)}{d\lambda}$ (see Fig.13a). However, we noticed that both direct and indirect band gap transitions occur at approximately the same energy for all BiFeO$_3$ nanoparticle samples, which could be due to their CBM band and Urbach tail band being situated at the same energy level. This observation requires further investigation to conclude. Additionally, a peak feature near 1.75 eV was seen due to one of the four expected d-d transitions for Fe$^{3+}$ ions in the octahedral FeO$_6$ of BiFeO$_3$.[42,43,83-85] These findings are consistent with our earlier observation using the F(R) data.[82]

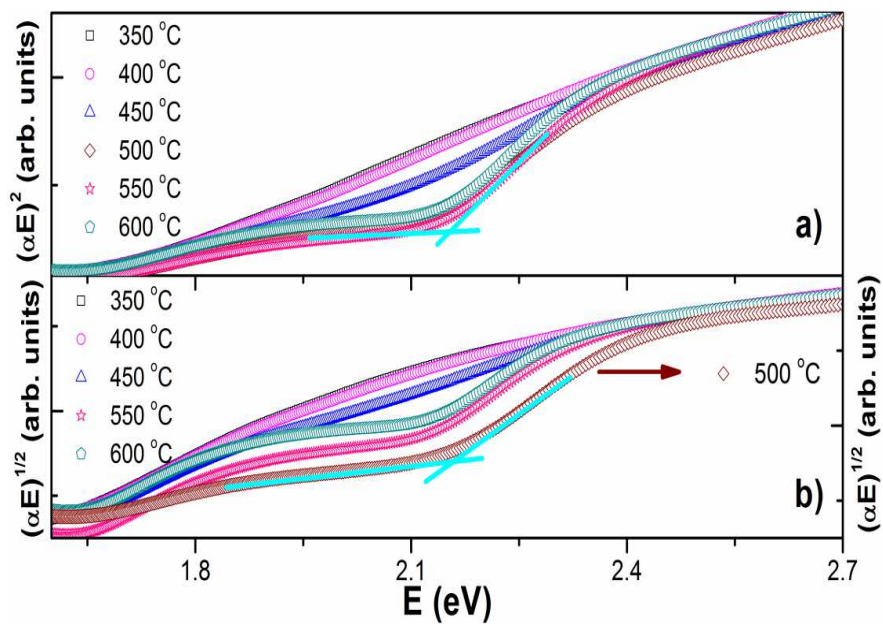

Figure 13 a) $(\alpha E)^2$ versus $E$ and b) $(\alpha E)^{1/2}$ versus $E$ for BiFeO$_3$ nanoparticle samples calcined at different temperatures. The magenta lines drawn correspond to the linear fitting for the estimation of $E_g$.

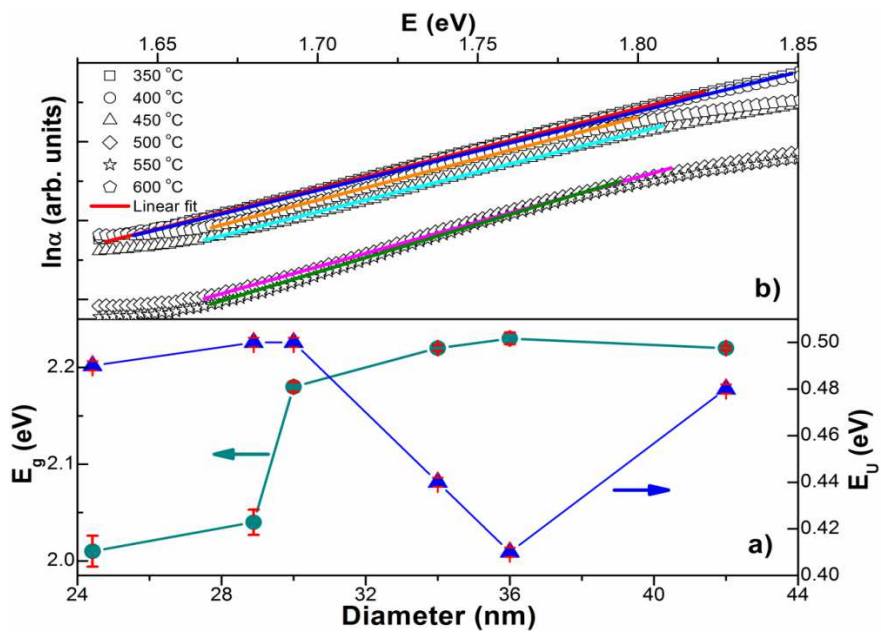

Figure 14 a) the band gap energy ($E_g$, on the left *y*-axis) and Urbach energy ($E_U$, on the right *y*-axis) versus the diameter (*D*) of nanoparticle samples and b) the fitted *lnα* versus *E* of the BiFeO$_3$ nanoparticle samples calcined at different temperatures. The solid lines presented in Fig. 14a serve as a visual guide, while the solid lines in Fig. 14b represent the best linear fit to the graphs of *lnα* versus *E*.

We also estimated the $E_U$ values of the BiFeO$_3$ nanoparticle samples using the linear regression method on their semi-logarithmic plots, $\ln\alpha$ versus $E$. The obtained $E_U$ values of these samples are plotted against their average diameter ($D$) on the right $y$-axis in Fig. 14b. It is found that the BiFeO$_3$ samples with small nanoparticles have high $E_U$ values (prepared at low temperatures, $T < 500$ °C) than the samples with large-sized nanoparticles that are prepared at high temperatures, $T \geq 500$ °C. This indicates the defect-induced Urbach tail band is deeper for the low-temperature heated BiFeO$_3$ nanoparticle samples with $E_U \sim 0.5$ eV than the high-temperature heated samples with $E_U < 0.5$ eV. This finding is in line with the observation using our structural analysis of these inspected samples.[82] The evaluated $E_U$ values of the BiFeO$_3$ nanoparticle samples are comparable to the literature values of the BiFeO$_3$-based nanoparticles.[71]

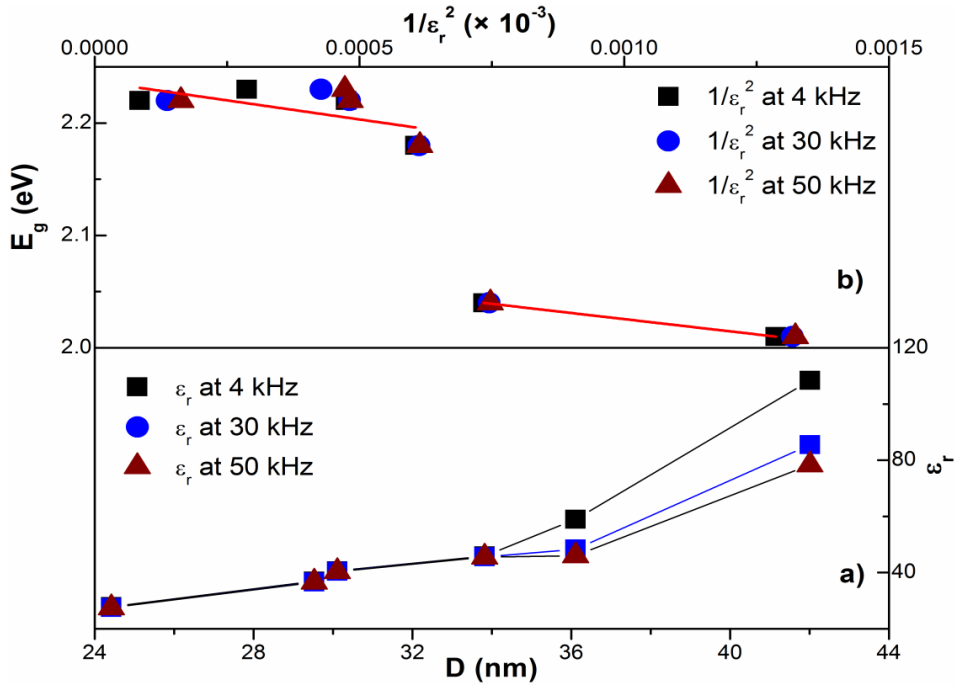

Figure 15 a) The dielectric constant ($\varepsilon_r$) of the BiFeO$_3$ nanoparticle samples measured at three different frequencies: 4 kHz, 30 kHz, and 50 kHz, and b) the plot of $E_g$ versus $\frac{1}{\varepsilon_r^2}$ for the BiFeO$_3$ nanoparticle samples. The solid lines in the figures are a visual guide for the eye.

The measured dielectric constant ($\varepsilon_r$) of the BiFeO$_3$ nanoparticle samples at three different frequencies of 4 kHz, 30 kHz, and 50 kHz is plotted against their diameter, as displayed in Fig. 15 a. The $\varepsilon_r$ values of these samples are similar to those of BiFeO$_3$-based nanoparticles reported in the literature.[83-85] Importantly, the $\varepsilon_r$ values for the BiFeO$_3$ nanoparticle samples were considerably higher than those of bulk BiFeO$_3$ (see Table 1).[74] Furthermore, we charted the estimated $E_g$ of the BiFeO$_3$ nanoparticles against the inverse of the square of their dielectric constant ($\frac{1}{\varepsilon_r^2}$) to analyze it using the Bohr-like model (i.e., $E_g = \frac{1}{\varepsilon_r^2}$) as specified by Rambadey *et al*.[75] It is clear that the changes in $E_g$ concerning $\varepsilon_r$ of the examined BiFeO$_3$ nanoparticle samples synthesized at elevated temperatures ($T > 400$ °C) align well with the Bohr-like model,[75] as the linear correlation of $E_g$ with the term $\frac{1}{\varepsilon_r^2}$ was demonstrated by roughly sketching the solid lines in Fig. 15b. The BiFeO$_3$ nanoparticle samples prepared at lower temperatures ($T \leq 400$ °C) also obey the Bohr-like model,[75] however they exhibit a different slope than the samples prepared at higher temperatures (Fig. 15b).

## G. Schematic band structure for the examined functional oxides and BiFeO$_3$

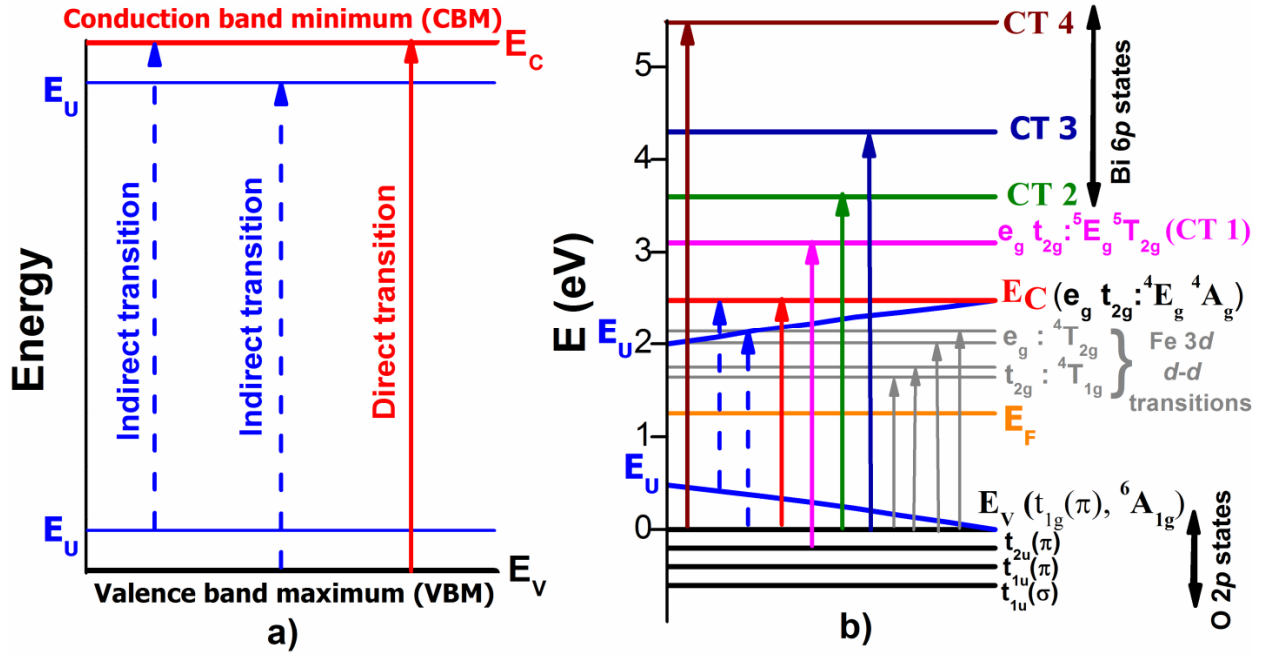

Figure 16 a) A representative band structure for the inspected functional oxides and b) electronic band structure of bulk BiFeO$_3$ based on the results obtained using diffuse reflectance spectroscopy and the KM function. The red, magenta, olive, royal blue, and pink solid arrows represent direct band gap and four charge transfer transitions, while the blue dashed arrows illustrate indirect interband transitions for the functional oxides. Four *d-d* transitions related to distorted Fe$^{3+}$O$_6$ octahedra in the BiFeO$_3$ are depicted as grey solid arrows.

To summarize our findings regarding the probed functional perovskite oxides using DRS spectroscopy coupled with the KM function and the Tauc plot analysis, we have developed a representative band structure as shown in Fig. 16a. This illustrates the mechanisms of different electronic transitions in these functional perovskite oxides. A direct band gap transition associated with the functional materials is indicated by a red-colored solid arrow. The detected Urbach energy resulting from the formation of the Urbach tail band caused by defects in the materials is illustrated by the blue-colored solid lines located beneath the conduction band and above the valence band. These potential indirect electronic transitions in the functional materials may occur from the CBM band to the Urbach band situated above the valence band and/or from the Urbach band positioned below the conduction band to the VBM band. The values obtained for direct band-gap energy and Urbach energy of the studied materials (except $BiMnO_3$) are listed in Table 2.

Table 2 The evaluated room-temperature values of direct band gap energy, Urbach energy, the sharpness parameter of the absorption edge, the electron-phonon energy, and the NEAR factor of the investigated functional materials (except $BiMnO_3$) using the relative reflectance measurement.

| Sample | $E_g$ (eV) | $E_U$ (eV) | $\beta(T)$ | $E_{e\text{-}ph}$ (eV) | NEAR factor = $\alpha(E_g)/\alpha(1.02E_g)$ |
|---|---|---|---|---|---|
| Bulk $V_2O_5$ | 2.34 | 0.24 | 0.108 | 6.17 | 0.86 |
| Bulk $BaSnO_3$ | 3.25 | 0.38 | 0.068 | 9.80 | 0.87 |
| Bulk $PbZr_{0.52}Ti_{0.48}O_3$ | 3.18 | 0.25 | 0.104 | 6.41 | 0.88 |
| Bulk $BiFeO_3$ | 2.40 | 0.48 | 0.054 | 12.35 | 0.96 |
| Nanostructured $BiFeO_3$ | 2.40 | 0.35 | 0.074 | 9.01 | 0.91 |
| $BiFeO_3$ NPs at 350 °C | 2.01 | 0.49 | 0.053 | 12.58 | 0.92 |
| $BiFeO_3$ NPs at 400 °C | 2.04 | 0.50 | 0.052 | 12.82 | 0.91 |
| $BiFeO_3$ NPs at 450 °C | 2.18 | 0.50 | 0.052 | 12.82 | 0.93 |
| $BiFeO_3$ NPs at 500 °C | 2.22 | 0.44 | 0.059 | 11.30 | 0.96 |
| $BiFeO_3$ NPs at 550 °C | 2.23 | 0.41 | 0.063 | 10.58 | 0.97 |
| $BiFeO_3$ NPs at 600 °C | 2.22 | 0.48 | 0.054 | 12.35 | 0.94 |

We also estimated the near-edge absorptivity ratio, NEAR factor for these functional materials, as developed by Viezbicke et al.[86-88] The NEAR factor for the direct allowed transition of the examined materials is deduced using a simple relation:

$$NEAR = \frac{\alpha(E_g)}{\alpha(1.02E_g)}. \quad (16)$$

The calculated NEAR factors of the materials are also listed in Table 2. We noticed that the NEAR factors of the BiFeO$_3$-based samples are slightly higher ($\geq 0.91$) than those of other studied materials ($\leq 0.88$). Interestingly, the $E_U$ values are larger for the BiFeO$_3$-based materials compared to other compounds. Thus, we argue that there is a significant influence of Urbach tail bands on the electronic structure of the inspected functional materials, given their NEAR factor ($\geq 0.86$) is close to 1 as well as their $E_U$ is also considerably large, about $\geq 240$ meV. On the contrary, the estimated temperature-dependent steepness parameter, $\beta(T)$, which describes the sharpness of the absorption edge is smaller for the BiFeO$_3$-based samples when compared to other examined perovskite oxides, by the relation: $E_U = \frac{K_B T}{\beta(T)}$.[89-91] The $\beta(T)$ values (~0.05–0.11) of all studied functional oxides are very low when compared to semiconductor In$_{1-x}$Ga$_x$As$_{1-y}$P$_y$ compounds (0.96–1.45).[89] However, their value is high when compared to the lead-free semiconductor compound CsMnBr$_3$ (~0.03).[90] Interestingly, we found that $\beta(T)$ values of our samples are comparable to that of the Mn-doped magnetoelectric GdFeO$_3$ materials (~0.02–0.06).[8] The current observations signify that the optical absorption edge in our perovskite materials is only moderately sharp. We also estimated the electron-phonon interaction energy ($E_{e-ph}$) using the relation:[90] $E_{e-ph} = \frac{2}{3\beta(T)}$, which is listed in Table 2. Particularly, we noted that the $E_{e-ph}$ energy for the BiFeO$_3$-based materials is in the range of ~9–12.8 eV, which is also comparable to the values (11.1–35.3 eV) of the GdFe$_{1-x}$Mn$_x$O$_3$ samples.[91]

We have also created a schematic band structure (as shown in Fig. 16b) for bulk BiFeO$_3$ as a representative band formation for the inspected BiFeO$_3$-based materials. This band diagram is proposed based on our comprehensive analysis of the diffuse reflectance data of all BiFeO$_3$-based samples using the relative reflectance and dilution methods with the KM ($F(R)$) function, a first derivative of reflectance ($\frac{dR(\lambda)}{d\lambda}$), and the Tauc plot analysis. From these detailed analyses, we evaluated the direct band gap energy to be about 2.5±0.1 eV. This transition is essentially due to a dipole-forbidden *p-d* charge transfer transition,[92-96] $t_{1g}(\pi) \to t_{2g}$ which is shown by a red solid arrow in Fig. 16b. While the dipole-allowed *p-d* charge transfer transition ($t_{2u}(\pi) \to t_{2g}$) is found at 3.0±0.1 eV,[80,96] which is displayed by a magenta solid arrow in Fig. 16b. Two other observed charge transfer transitions near 3.5±0.1 eV and 4.3±0.1 eV belong to *p-p* charge transfer transitions (CT 2 and CT 3 are depicted by olive and royal blue solid arrows in Fig. 16b) between Bi 6*p* and O 2*p* states of BiFeO$_3$.[82,92,93] Additionally, to mark the Fermi energy ($E_F$, see a orange solid line) and higher-energy charge transfer (CT 4 as shown by a wine solid line) transition in our proposed band structure, we also plotted the energy separation between the valence band and the Fermi energy ($E_V$-$E_F$ = 1.25 eV) and higher-energy Bi 6*p* states near 5.5 eV of bulk BiFeO$_3$ that were obtained using X-ray photoelectron spectroscopy study.[42,97]

Two doubly degenerate *d-d* transitions in the BiFeO$_3$-based systems with Fe$^{3+}$O$_6$ octahedra were detected near 2.1, 2.0, 1.8, and 1.6 eV,[42,43] resulting from the creation of electron [FeO$_6$]$^{10-}$ and hole [FeO$_6$]$^{8-}$ related to Fe$^{2+}$ and Fe$^{4+}$ ions upon incident photons.[92-94] These transitions are assigned as e$_g \to$ e$_g$, t$_{2g} \to$ e$_g$, t$_{2g} \to$ t$_{2g}$, and e$_g \to$ t$_{2g}$,[92] and are illustrated by grey arrows in Fig. 17b. Two possible indirect transitions with the energy of 2.1±0.1 eV due to induced defects are shown by the blue dashed arrows in Fig. 17b. These indirect transitions typically occur from the CBM to the Urbach band above the valence band and/or from the Urbach band below the conduction band to the VBM. This is attributed to the observed Urbach tail band with the energy of $E_U$ = 0.5±0.1 eV that is commonly expected to be located below the CBM and/or the VBM of bulk BiFeO$_3$. It was also observed that the energies of optical band gap and charge transfer transitions in multiferroic BiFeO$_3$ are slightly altered with (Ca, Ba)-doping and size reduction in the BiFeO$_3$ nanoparticle samples when compared to the bulk BiFeO$_3$.[35,42,43,82,97] Particularly, a gradual reduction in the band gap energy with a decrease in the *c/a* ratio (as chemical pressure increases) was achieved via (Ba, Ca)-doping in BiFeO$_3$[43,97] and reduction in particle size of the BiFeO$_3$ nanoparticles,[82] which is illustrated in Fig. 17. In conclusion, we have shown that the optical properties including the optical band gap of the functional material like BiFeO$_3$ can be tuned by induced chemical pressure via suitable doping and size reduction to achieve desirable properties for many optical applications including photoferroelectric,[29] photocatalytic,[39] and photovoltaic device applications.[98]

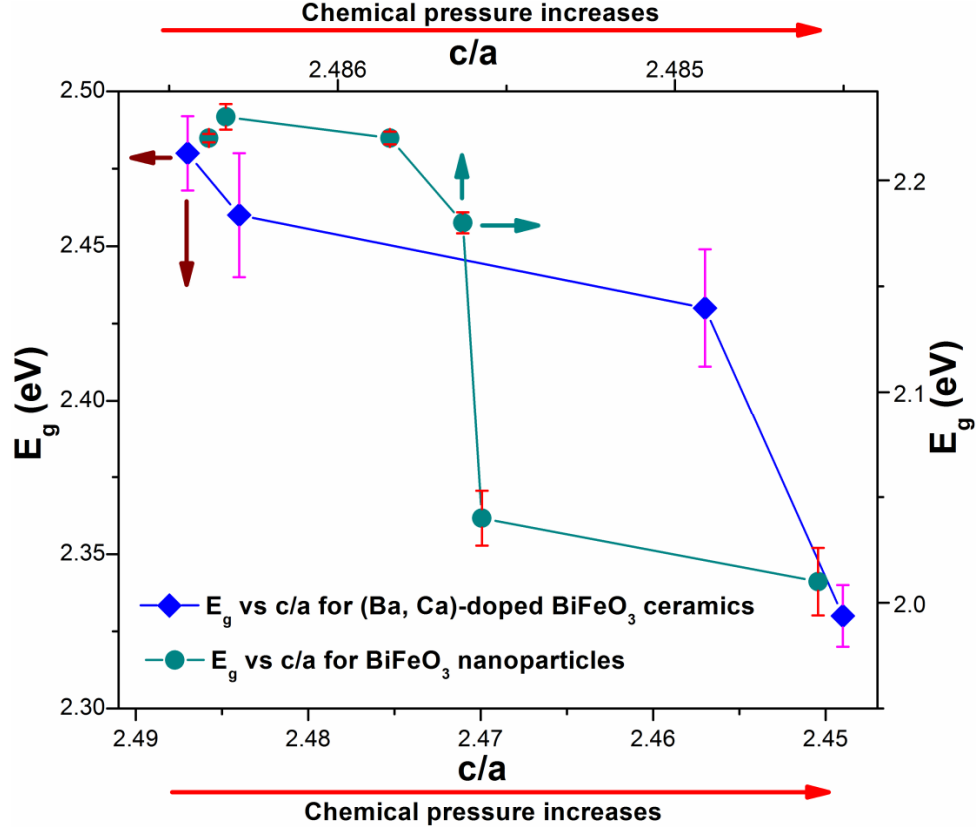

Figure 17 The band gap energy ($E_g$) versus $c/a$ ratio for the (Ba, Ca)-doped BiFeO$_3$ samples (see the bottom $x$-axis and the left $y$-axis) and BiFeO$_3$ nanoparticle samples (see the top $x$-axis and the right $y$-axis). The solid lines serve as guides for the eye, and the errors in the band gap energy of the samples are provided as standard deviations.

## IV. CONCLUSION

The optical properties of various functional perovskite oxides were investigated by studying their electronic energy levels and defects-induced Urbach tail bands through diffuse reflectance spectroscopy. In particular, the magnetoelectric multiferroic BiFeO$_3$ in both its bulk and nano forms has been studied to investigate its electronic transitions, including charge transfer and $d$-$d$ transitions. This was accomplished using diffuse reflectance obtained via the dilution method and the relative reflectance technique. The first derivative of reflectance, the Tauc plot, and the Kubelka-Munk function are employed to determine the energies of direct and indirect electronic transitions for each perovskite oxide. The direct band gap energy ($E_g$) of bulk V$_2$O$_5$, BaSnO$_3$, PbZr$_{0.52}$Ti$_{0.48}$O$_3$, and BiFeO$_3$ materials was estimated to be approximately 2.27, 3.25, 3.1, and 2.48 eV, respectively, consistent with the existing literature. Furthermore, we evaluated the defects-induced Urbach energy ($E_U$) values for the direct band gap functional oxides to be approximately 0.24, 0.38, 0.25, and 0.48 eV, respectively, by studying the exponential behavior near their optical absorption onsets. A noteworthy finding is that the band gap energy of $BiFeO_3$ gradually decreases with an increase in chemical pressure from (Ba, Ca)-doping into the host lattice and with a reduction in the size of the $BiFeO_3$ nanoparticles. Considering the potential of the model multiferroic compound $BiFeO_3$ for numerous applications including optoelectronic devices, we provided a detailed electronic band structure for $BiFeO_3$ and outlined a representative band structure for all functional perovskite oxide materials under investigation.

**SUPPLEMENTARY MATERIAL**

See the supporting material for detailed information on the X-ray diffraction pattern of the prepared $V_2O_5$ and sintered PZT(52/48) samples, the estimated relative error in the Kubelka-Munk (KM) function versus reflectance for the functional oxides, the plot of the first derivative of reflectance for a commercial $V_2O_5$ sample, the spectra of the KM function and absorption coefficient for the synthesized $α$-$V_2O_5$ and commercial $V_2O_5$ samples, the reflectance and KM function plots for bulk $BaSnO_3$, the KM function data and the first derivative of reflectance for bulk PZT(52/48). Additionally, refer to the supporting material for the KM function versus energy for $BiFeO_3$ and $BiMnO_3$ samples, the first derivative of reflectance data for pure and doped $BiFeO_3$ samples, the X-ray diffraction data for the synthesized $BiFeO_3$ nanostructured sample, and the deconvoluted first derivative reflectance data for $BiFeO_3$ nanoparticle samples.

**AUTHOR DECLARATIONS**

**Conflict of interest**

The authors have no conflicts to declare.

**Author contributions**

**Ramachandran Balakrishnan:** Conceptualization, Methodology, Data Curation (lead), Formal Analysis (lead), Writing - Original Draft, Writing - Review & Editing (equal), Project Administration (lead). **Priyambada Sahoo:** Data Curation (supporting), Formal Analysis (supporting), Writing - Review & Editing (equal). **Balamurugan Karuppannan**: Data Curation (supporting), Formal Analysis (supporting), Writing - Review & Editing (equal). **Ambesh Dixit**: Formal Analysis (supporting), Writing - Review & Editing (equal), Funding Acquisition, Project Administration (supporting).

## DATA AVAILABILITY

The authors declare that the data that supports this work are available within the article and its supporting material.

## ACKNOWLEDGMENT

Ambesh Dixit acknowledges Defence Research and Development Organization (DRDO), New Delhi through project # ERIP/ER/202401002/M/01/1857 for carrying out this work.## REFERENCES